%% file: main.tex
\documentclass{article}
\usepackage{graphics}
\usepackage[english]{babel}
\usepackage[utf8x]{inputenc}
\usepackage[per-mode=symbol]{siunitx}
\usepackage{amsmath}
\usepackage{amssymb}

\usepackage{authblk}
\usepackage{graphicx}
\usepackage{dcolumn}
\usepackage{bm}
\usepackage{vmargin}
\setpapersize{A4}
\setmargins{2.5cm}
{1.5cm}                        
{16.5cm}                      
{23.42cm}                    
{10pt}                           
{1cm}                           
{0pt}                             
{2cm}

\usepackage{graphicx}
\usepackage{dcolumn}
\usepackage{bm}

\usepackage{textcomp}

\title{Area covered by disks in small-bounded continuum percolating systems: An application to the string percolation model}

\author[1,2]{J. E. Ram\'irez\thanks{jerc.fis@gmail.com}}
\author[1,2]{C. Pajares\thanks{pajares@fpaxp1.usc.es}}
\affil[1]{Departamento de F\'isica de Part\'iculas, Universidad de Santiago de Compostela, E-15782 Santiago de Compostela, Espa\~na}
\affil[2]{
Instituto Galego de F\'isica de Altas Enerx\'ias, Universidad de Santiago de Compostela, E-15782 Santiago de Compostela, Espa\~na}

\date{}

\begin{document}
\maketitle

\begin{abstract}
In string percolation model, the study of colliding systems at high energies is based on a continuum percolation  theory in two dimensions where the number of strings distributed in the surface of interest is strongly determined by the size and the energy of the colliding particles.
It is also expected that the surface where the disks are lying be finite, defining a system without periodic boundary conditions.
In this work, we report modifications to the fraction of the area covered by disks in continuum percolating systems due to a finite number of disks and bounded by different geometries: circle, ellipse, triangle, square and pentagon, which correspond to the first Fourier modes of the shape fluctuation of the initial state after the particle collision.
We find that the deviation of the fraction of area covered by disks from its corresponding value in the thermodynamic limit satisfies a universal behavior, where the free parameters depend on the density profile, number of disks and the shape of the boundary.
Consequently, it is also found that the color suppression factor of the string percolation model is modified by a damping function related to the small-bounded effects. 
Corrections to the temperature and the speed of sound defined in string systems are also shown for small and elliptically bounded systems.
\end{abstract}



\section{Introduction}

The string percolation model is an alternative description of the collective behavior observed in nucleus-nucleus collisions due to the formation of quarks and gluon matter which subsequently expands as a liquid with very small shear viscosity over entropy ratio  \cite{FOKA2016154, GYULASSY200530,SHURYAK200948}.

In string percolation multiparticle production is described in terms of color strings stretched between the partons of the projectile and target. This strings decays into $q-\bar{q}$ and $qq-\bar{q}\bar{q}$ pairs and subsequently hadronize producing the observed hadrons. Due to confinement, the color of these strings is confined to a small area, $a_0$ in the transverse plane.
With increasing energy and/or size and centrality of the colliding objects, the number of strings grows and the strings start to overlap forming clusters, very similar to clusters of disks in two-dimensional continuum percolation theory.
At a critical density, a macroscopic cluster appears crossing the transverse collision surface. The clusters of strings behave similarly to a single string with a higher color field corresponding to add the individual field of each string, implying a higher tension and thus a large mean transverse momentum \cite{Braun2015,RMF,Pajares2005,DIASDEDEUS,Braun2002,PhysRevC.69.034901,PRCBraun}.
In the applications of string percolation to describe RHIC and LHC data on AA collisions, plays an important role the fraction of area covered by disks, which in the thermodynamic limit is given by \cite{Kertez,Braun2000}
\begin{equation}
    \phi_\text{TL}(\rho)=1-\exp(-\rho),
\label{eq:CAD-TL}
\end{equation}
where $\rho$ is the filling factor, defined by $\rho=Na_0/S$, where $N$ is the number of disk, $a_0$ the area of one disks and $S$ the transverse area where the disks are distributed.
The number of disks (strings) formed in a collision depends on the profile of the projectile and target, energy and impact parameter (centrality degree) of the collision. In heavy-ion collisions as Au-Au collisions at RHIC energies or Pb-Pb collisions at the LHC energies, $N$ is very large, close to two thousands in Pb-Pb collisions and it is expected to be close to the thermodynamical limit in a percolation approach.
However, this is not longer true in te case of pA or pp collisions, where $N$ is only a few tens for a minimum bias collisions \cite{IBPRD,BAUTISTA2012230}.

The experimental data on the azimuthal distributions of the momenta of the produced particles and the corresponding Fourier coefficients, $v_n$, show a collective behavior in pp and pA collisions very similar to the previously observed in Au-Au and Pb-Pb collisions \cite{PhysRevLett.107.032301,Chatrchyan2014,Alice1,Khachatryan2010,Jacak310,Alice1,Alice2,Alice3,Alice4,Alice5,Qin}. The studies of the collective behavior and the different harmonics in string percolation usually assume to be close to the thermodynamic limit \cite{IBPRD,BAUTISTA2012230,Bautista2012,Braun2018}, but it is well-known that the properties of percolating systems (percolation threshold, cluster density) may be modified by topology, system size and periodic boundary conditions \cite{Domb,PhysRevLett.75.193,CHINKUNHU1995,SABERI2015}. 
In this paper, we show the modifications to the fraction of area covered by disks, due to the small number of disks for systems free of periodic boundary conditions where the surface $S$ takes the form of circle, ellipse, square, pentagon,$\dots$ corresponding to the different space eccentricities $e_1$, $e_2$, $e_3$, $e_4$, $e_5$, $\dots$ The modification leads to corrections to the color suppression factor $F(\rho)$ which is relevant in string percolation for describing transverse momentum and multiplicity distributions.

The underlying physical grounds of the collective behavior produced in the collisions of small systems like pp are under debate, in particular whether it is originated by initial state effects or on the contrary by final state interactions suitable to a hydrodynamical description. In this context it is important to know whether the collective behavior occurs also for rather low multiplicity in pp collisions.
In both cases, it is crucial to know the profile of the proton and its fluctuations to describe the experimental data. In this line, it has been emphasized the proton shape fluctuations in the explanation of the incoherent diffractive vector meson production \cite{PhysRevLett.117.052301, CEPILA2017186}, the importance of the core-corona proton profile to explain the collective behavior seen in pp collisions \cite{PhysRevC.89.064903} and the rather large edge and the existence of correlated hot spots inside of the proton to explain the energy evolution of the scattering amplitude of elastic pp collisions \cite{ALBACETE2017149}.

In this paper, we also study the  effects of different profiles in pp collisions in the framework of string percolation
and the differences between the evaluated observables in the thermodynamical limit and in the case of rather small degrees of freedom (string) in pp collisions.

The paper is organized as follows. In Section~\ref{model}, we describe the model to generate bounded string percolating systems for a small number of disks. We also present the algorithms to fill the system with three different density profiles and the computation of the fraction of area covered by disks.
In Section~\ref{results}, we report the modification of the fraction of area covered by disks as a function of the number of disks, boundary shape and density profile. We also diskuss the limit of high densities for small-bounded systems. Section~\ref{APP} shows the modifications to the color suppression factor as a second damping factor to the color field of the strings due to small-bounded string systems and the analysis of the critical behavior of the speed of sound around the critical temperature in small-bounded string percolating systems. Finally, Section~\ref{Conclusions} contains the conclusions of this work.

\section{Model}\label{model}

In this section, we present the algorithms to generate string systems bounded by circles, ellipses, triangles, squares, and pentagons, considering different density profiles. It is also presented the corresponding algorithm to determine the fraction of area covered by disks in such systems.

\subsection{Construction of geometric boundaries}\label{boundary}

The continuum percolation systems are defined as a collection of randomly distributed and fully penetrable objects (disks, ellipses, rotated squares, sticks, for example see Refs.~\cite{RSA:RSA20064,PhysRevE.66.046136,PREMertens}) on a flat surface $S$ with periodic boundary conditions. Nevertheless, the requirements to define string percolating systems (string number and the overlapping area) are addressed by the parameters in high energy collisions: Size of colliding particles, energy at the center of mass, impact parameter, etc. In particular, the string percolating system is formed only by disks lying in the overlapping area, being necessary a description of two-dimensional continuum percolation without periodic boundary conditions.

A general description of two-dimensional continuum percolation is given through the filling factor.
In a system with fixed both the number of disks and the shape of the boundary, one way to get variations in the filling factor is changing the surface of the overlapping area. In this way the dimensions of  $S$ can be written as a function of $\rho$, $N$, and the geometry information. 
For a boundary shape defined by a $n$-sides regular polygon, the length of the sides is given by
\begin{equation}
\rho=\frac{4N\pi r_0^2 \tan(\pi/n)}{nl^2},
\end{equation}
where $r_0$ is the radio of the small disk and $l$ is the length of the sides of the regular polygon. Note that in the above expression, for a given number of disks $N$ and density one can determine the dimension of the regular polygon of $n$ sides which confines the system by the following relation
\begin{equation}
l=\sqrt{\frac{4N\pi \tan(\pi/n)}{n\rho}}r_0.
\label{eq:l-pol}
\end{equation}
In the limit $n\to \infty$, we found that the expression in Eq.~\eqref{eq:l-pol} takes the form $l\to 0$, which is in agreement with the boundary circle limit, i. e., in this case an $n$-sides regular polygon looks like a circle. Moreover, the semi-axes for an elliptic boundary are given by
\begin{eqnarray}
A^2=\frac{r_0^2N}{\rho \sqrt{1-\varepsilon^2}},& &B^2=\frac{r_0^2N\sqrt{1-\varepsilon^2}}{\rho},
\label{eq:ab}
\end{eqnarray}
where $A$ and $B$ are the major and the minor semi-axes, respectively, and $\varepsilon$ is the eccentricity (this case has been recently diskussed  in Ref.~\cite{jerc}). The particular case $\varepsilon=0$ corresponds to a circular boundary, which is the classical continuum percolation model bounded by a circle, already studied by several authors \cite{SATZ20033c,PRCBraun,AMELIN1993312,CELIK1980128}.

Once determined the dimensions of the confinement surface, we take $N$ random points distributed according to a  particular density profile, representing the center of the disks and those should be at least a distance $r_0$ from any border, so the disks must be embedded entirely in the overlapping area.

\subsection{Density profiles}\label{densityP}

\begin{figure}[ht]
\centering
\includegraphics[scale=0.2]{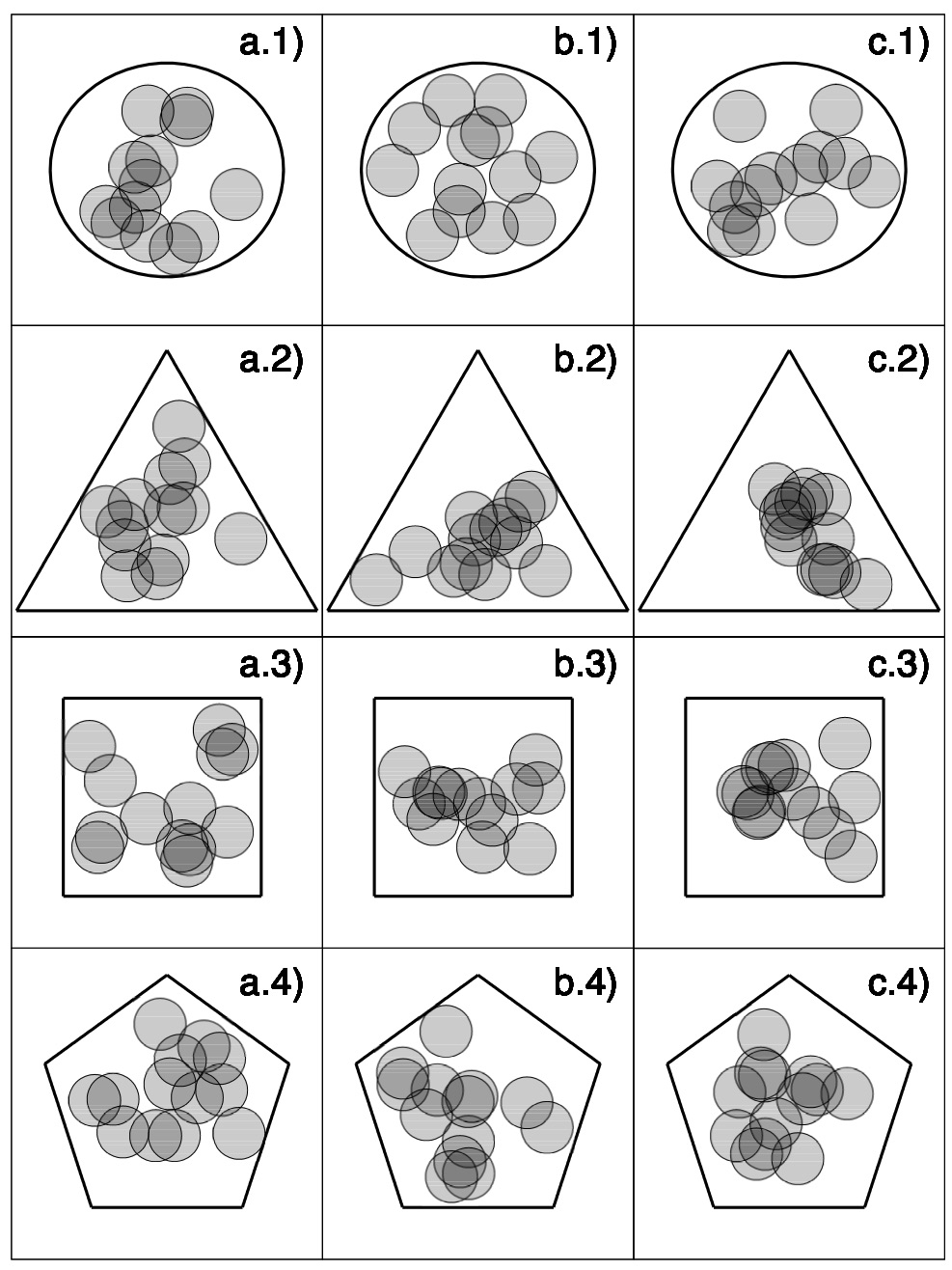}
\caption{Samples of string percolating system at $\rho=0.7$ for small number of disks ($N=13$), different density profile (columns) and boundary shape condition (rows).}
\label{fig:per-sample}
\end{figure}

\begin{figure}[ht]
\centering
\includegraphics[scale=0.2]{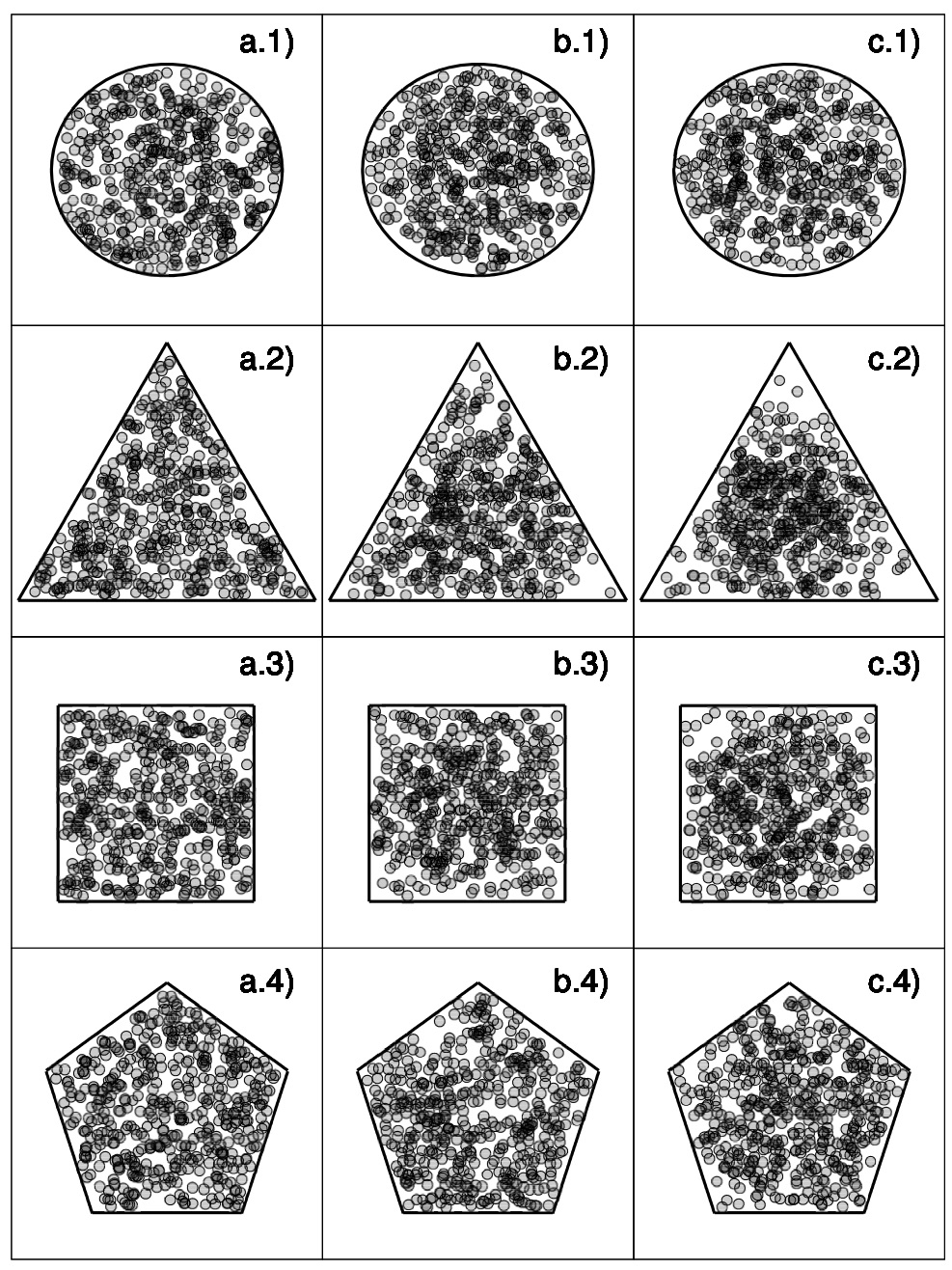}
\caption{Samples of high populated disk percolating system ($N=500$) at $\rho=1.1$ for different density profile (columns) and boundary shape condition (rows).}
\label{fig:per-sample2}
\end{figure}

Usually, in the studies of continuum percolation the considered disks are uniformly distributed.
However, the nuclear profiles function considered in heavy-ion collisions are more realistic. 
The most important feature of these density profiles is that they are denser in the center of the overlapping area and more dilute as we go away from the central region \cite{RODRIGUES1999402,Amelin2001}. One way to simulate the density profile is generating random points according to a Gaussian distribution function, defined as follow:
\begin{equation}
f(x,y)=\frac{1}{2\pi\sigma_x\sigma_y}\exp\left[-\frac{1}{2}\left( \frac{(x-x_0)^2}{\sigma_x^2}+ \frac{(y-y_0)^2}{\sigma_y^2}\right)  \right],
\end{equation}
where the vector $(x_0,y_0)$ is the center of the geometric boundary shape and we have considered non-correlated distribution over the axes $x$ and $y$.
Note that there is not a unique way to define the values of the standard deviation. For regular polygons we choice $\sigma_x=\sigma_y=\sigma$, taking the following values
\begin{subequations}
\begin{eqnarray}
\sigma &=& R_{I},\label{1s2}\\
\sigma &=& R_{I}/2^{1/2},\label{sq2}
\end{eqnarray}
\end{subequations}
where $R_{I}$ is the radius of the inscribed circle in the bounded region. However, for the elliptical shape, we use different dispersion along the semi-axes: 
\begin{subequations}
\begin{eqnarray}
\sigma_x=(A-r_0),&\sigma_y=(B-r_0), \label{1s}\\
\sigma_x=(A-r_0)/2^{1/2},&\sigma_y=(B-r_0)/2^{1/2}. \label{sq}
\end{eqnarray}
\end{subequations}
We denote by 1S and SQ the profiles defined by the standard deviation ~\eqref{1s2}-\eqref{1s} and \eqref{sq2}-\eqref{sq} respectively. The case of disks uniformly distributed is denoted by U. The cases U and SQ in the thermodynamical limit have been used by different groups to study high energy colliding systems. Figures~\ref{fig:per-sample} and~\ref{fig:per-sample2} show examples of the three different cases.

\subsection{Measurement of the area covered by disks}

\begin{figure}
\centering
\includegraphics[scale=0.2]{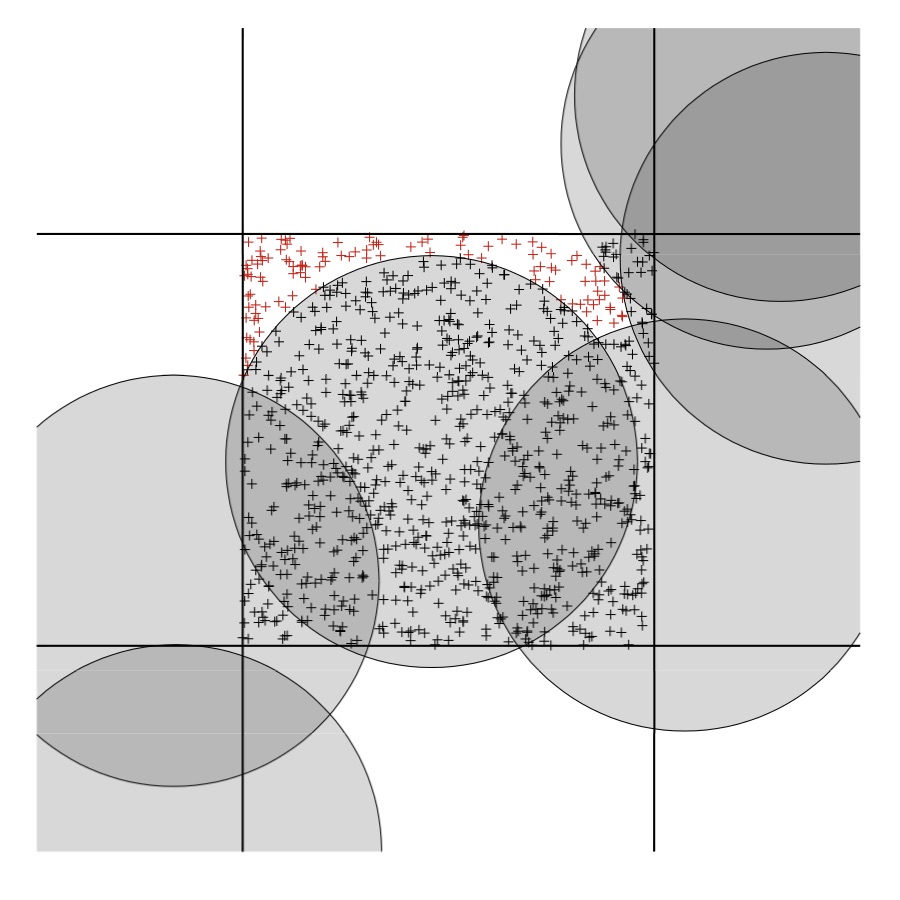}
\caption{Scheme of the Monte Carlo integration by cell to measure the fraction of area covered by disks. We count only points lying in disks on the cell or any of the nearest and next-to-nearest cells (black crosses).}
\label{fig:mcad}
\end{figure}

In order to determine the fraction of area covered by disks in string systems, we generate percolating systems according to the boundary and density profiles described in the Sec. \ref{boundary} and Sec. \ref{densityP}. 
Below we draw an imaginary square lattice with spacing $2r_0$ centered in the geometric center of the boundary, which allows us to map the string system into a matrix.
The disk centers lying in the $i$, $j$ cell are stored in the $i$, $j$ entry.
We then determine the fraction of the area covered by disks in each cell by Monte Carlo integration.
To do this, we generate random pairs $(x,y)$ uniformly distributed in the cell and counting the number points whose distance to any disks be less than $r_0$ in the cell or the nearest and the next-to-nearest-neighbor cells, as is depicted in Fig. 3, where the black crosses are the points lying in a disk (and red crosses are not).
This process is repeated for all cell in the lattice.
Finally, the total fraction of area covered by disks is determined as $4r_0^2\mathcal{N}_\text{in} /\mathcal{N}S$, where $\mathcal{N}_\text{in}$ is the total number of points laying in a disk of the total  $\mathcal{N}$ generated points.

\section{Results}\label{results}

In this section, we present the simulation result about the fraction of the area covered by disks for small-bounded string systems for the density profiles described in Sec. \ref{model}.

\begin{figure*}
\centering
\includegraphics[scale=0.2]{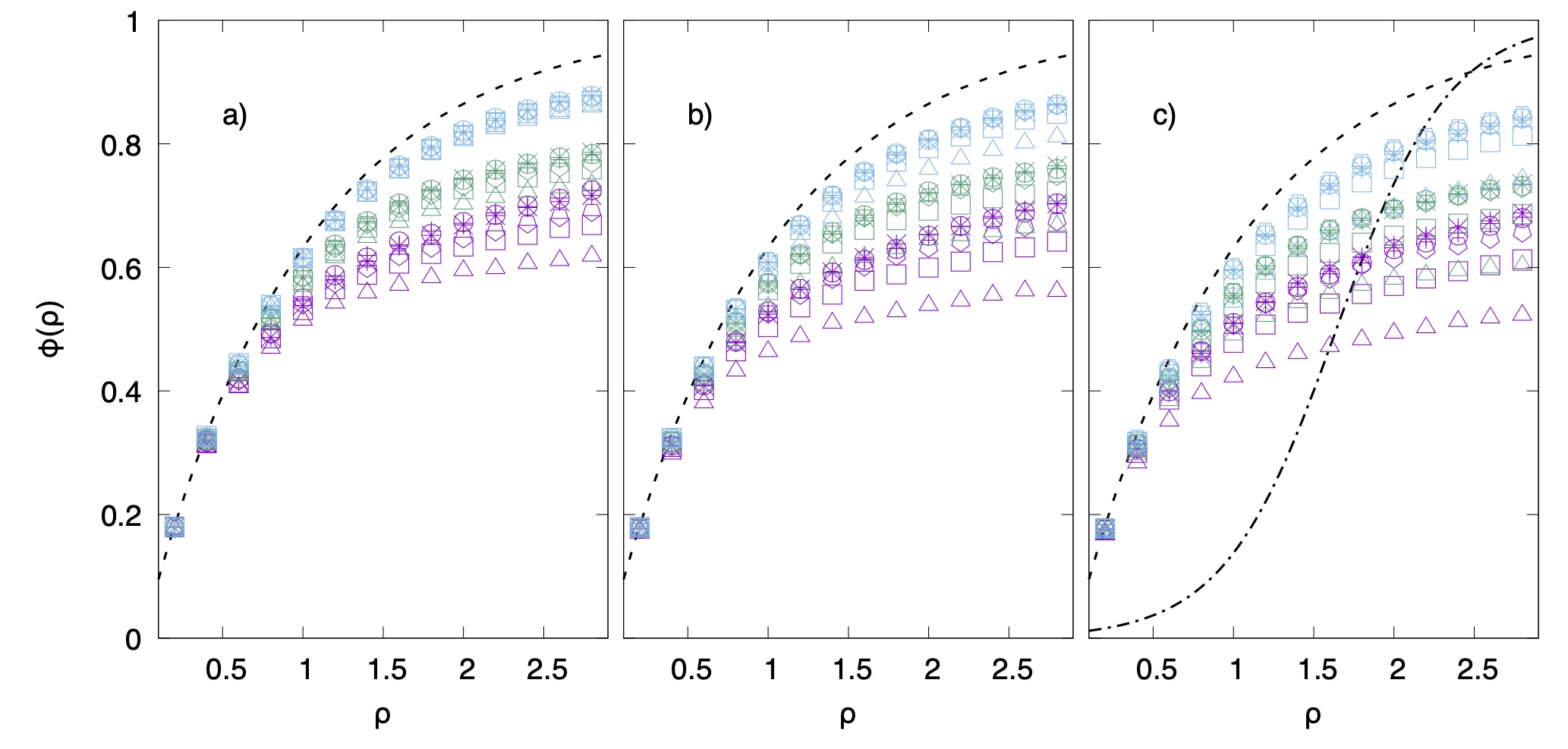}
\caption{Fraction of the area covered by disks in small continuum percolating systems for $N=$13 (purple), 96 (green) and 500 (cyan), and bounded by circles (circles), ellipses (crosses and stars), triangles (triangles), squares (squares) and pentagons (pentagons). The systems where filled according to three different density profiles: a) U, b) 1S and c) SQ. Dashed lines and dot-dash line are the fraction of area covered by disks in the thermodynamic limit for the U and SQ density profiles, respectively.}
\label{fig:f}
\end{figure*}

The fraction of area covered by disks is determined by averaging the area covered by disks over $2\times 10^5$ realizations at $N$, $\rho$, and boundary shape fixed. Additionally, the Monte Carlo integration was performed generating $10^4$ points per cell.
For each value of $N$=13, 55, 96 and 500, and each geometry, we measured the fraction of area covered by disks as a function of the filling factor, starting in $\rho=0.2$ until $\rho=2.8$, with increments of $\Delta \rho =0.2$.
This procedure is repeated for all three density profiles.
In Fig.~\ref{fig:f}, we show the explicit dependence of the fraction of area covered by disks on the number of disks, the boundary shape, and the density profile for string systems. 
Note that the fraction of area covered by disks in the uniform density profile recovers its behavior at the thermodynamic limit as $N$ increases and it becomes independent on the boundary shape, as is shown in Fig.~\ref{fig:f} a).
This is expected because in the thermodynamic limit there is no difference between large circles and ellipses or regular polygons.

\begin{figure*}
\centering
\includegraphics[scale=0.2]{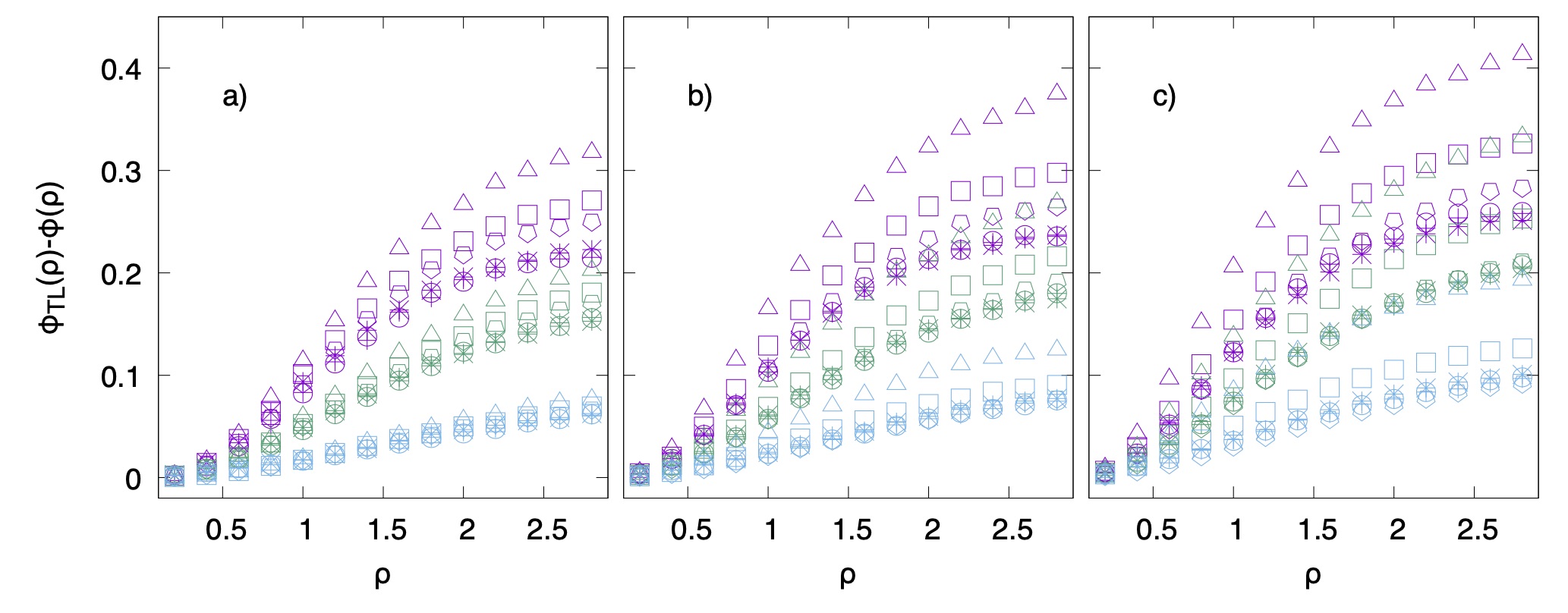}
\caption{Deviation of the fraction of area covered by disks in small-bounded system from its value in the thermodynamic limit for the uniform density profile for systems with $N=$13 (purple), 96 (green) and 500 (cyan), and bounded by circles (circles), ellipses (crosses and stars), triangles (triangles), squares (squares) and pentagons (pentagons), and considering different density profiles: a) U, b) 1S and c) SQ.}
\label{fig:df}
\end{figure*}

In order to analyze the data, it is convenient to define the deviation of the fraction of the area covered by disks from the thermodynamic limit
\begin{equation}
    \Delta \phi(\rho)=\phi_\text{TL}(\rho)-\phi(\rho),
\end{equation}
where $\phi_\text{TL}$ is the thermodynamical limit value given by Eq.~\eqref{eq:CAD-TL} and $\phi$ is the evaluated value by simulation.
Note that $\Delta \phi$ takes a sigmoid form and from $\phi$, it inherits all dependencies on the number of disks, boundary shape, and density profile, as we shown in Fig~\ref{fig:df}. The data of $\Delta \phi$ is well fitted by
\begin{equation}
\Delta \phi(\rho)=a\left[ 1+\tanh \left( \frac{\rho-\rho^*}{\Delta L}  \right)  \right],
\label{eq:df}
\end{equation}
where $a$ corresponds to the value where $\Delta \phi$ reaches a plateau behavior, $\Delta L$ is the width of the sigmoid transition, and $\rho^*$ is the filling factor value where $\Delta\phi(\rho^*)=a/2$.
In this way, we found the modified fraction of area covered by disks in small-bounded systems can be written as
\begin{equation}
\phi(\rho)=1-\exp(-\rho)-a\left[ 1+\tanh \left( \frac{\rho-\rho^*}{\Delta L}  \right)  \right],
\label{eq:f-cor}
\end{equation}
where the fit parameters $a$, $\Delta L$ and $\rho^*$ take particular values for each number of disks, boundary shape and density profile, which are summarized in Table~\ref{tab:pfits}.
It is necessary to remark that in the range of analyzed values of the filling factor we do not found a substantial difference in the fraction of area covered by disks between circles and ellipses independently on the number of disks and density profile. This last fact it is clearly appreciate in Fig.~\ref{fig:f} and Fig.~\ref{fig:df}.
As is expected, $\Delta \phi$ vanishes as $N$ increases, which naturally corresponds to the fraction of area covered by disks in the thermodynamic limit for the uniform density profile.
It is worth to mention the fraction of area covered by disks in the Gaussian density profile significatively deviates from its value in the thermodynamic limit, however, for very populated systems ($N\sim 5000$), that functional structure is (dash-dot line in Fig.~\ref{fig:f} c)), given by
\begin{equation}
    \phi^\text{Gauss}_\text{TL}(\rho)=\frac{1}{1+a^\text{Gauss}\exp(-(\rho-\rho_c)/b^\text{Gauss})},
\label{eq:CAD-sq}
\end{equation}
where $\rho_c\sim 1.5$ is the percolation threshold for the Gaussian density profile \cite{RODRIGUES1999402}.
The parameters $a^\text{Gauss}=1.5$ and $b^\text{Gauss}$ (0.35 for pp and pA collisions, and 0.75 for AA collisions) depend on the profile function \cite{Andrs2014}. In particular $b^\text{Gauss}$ controls the ratio between the width of the border of the profile and the total area \cite{Braun2015}.

\input{tabla}

\begin{figure}
\centering
\includegraphics[scale=0.2]{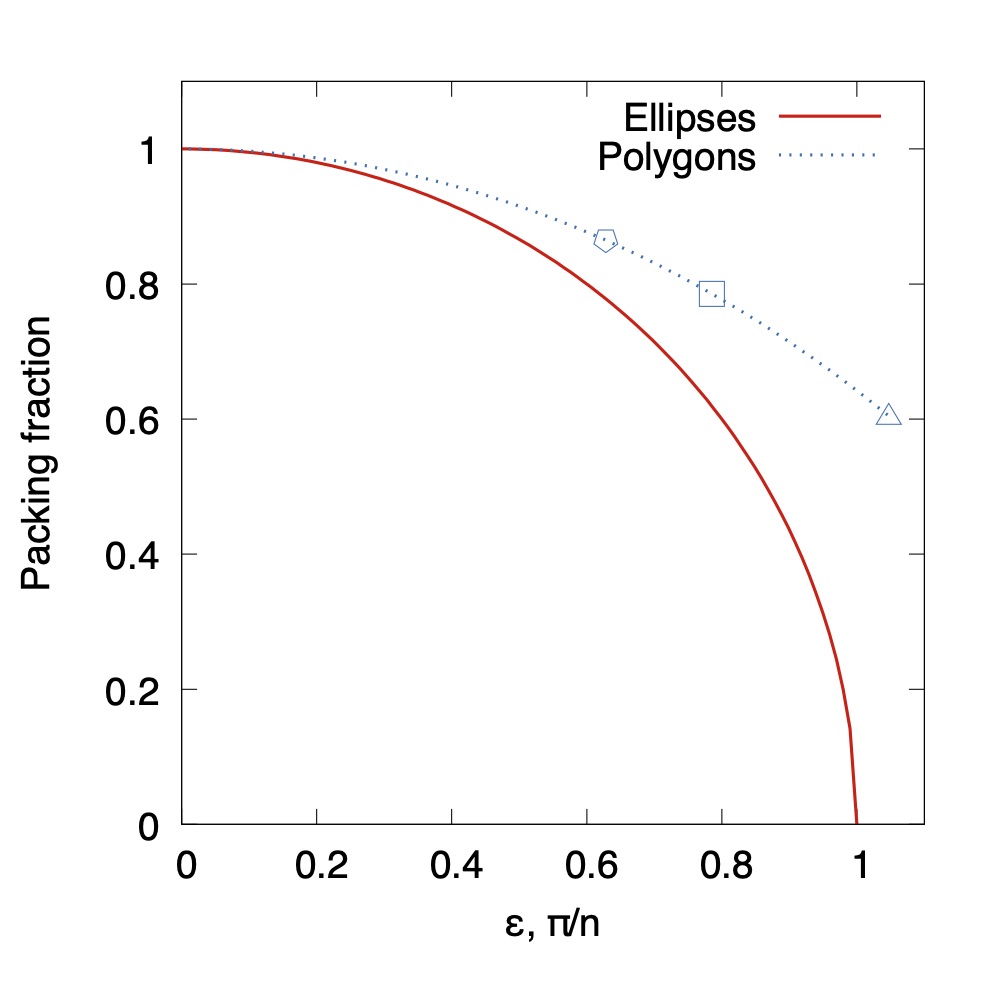}
\caption{Fraction of the area covered by disks of small-bounded continuum percolating systems at $\rho_\text{max}$ for ellipses (red line) as a function of the eccentricity and polygons (blue dotted line) as a function of $\pi/n$.}
\label{fig:pack}
\end{figure}

Moreover, for a percolating system with $N$ fixed,  there is a maximum value of filling factor that the system can reach. This occurs when the boundary is small enough to concentrate all disks center on the same point, which coincides with the geometrical center of the boundary.
At this point, the filling factor is
\begin{equation}
\rho_\text{max}=\left\{ \begin{array}{cc}
             N\sqrt{1-\varepsilon^2} &   \text{for ellipses,}  \\
              \frac{\pi N}{n\tan (\pi/n)} &  \text{for polygons.} 
             \end{array}
   \right.
\end{equation}
When the percolating system is filled at $\rho_\text{max}$, both the minor axis of an ellipse or the apothem of a polygon takes the value $r_0$.
Then, in the limit of high densities, the fraction of area covered by disks reach an asymptotic value, which corresponds to the packing fraction given by $\rho_\text{max}/N\leq1$. 
Figure \ref{fig:pack} shows the packing fraction as a function of $\varepsilon$ and $\pi/n$ for ellipses and polygons, respectively. Note that in the limit  $\varepsilon\to 0$ and $n\to\infty$, the fraction of area covered by disks goes to one, which is in full agreement with the circular boundary.

\section{Application to string percolation model}\label{APP}

In the string percolation model, the number of string and therefore the filling factor of its percolating picture depends strongly on the nature of the particles in the collision as well as the energy in the center of mass. For example, in pp collisions,  a rise in the string number is expected as the energy of the center of mass increase. Then, the disks start to overlap forming clusters. According to the summation rules of the color field, the color of the cluster is different from those single strings. 
In the following subsection, we estimate the modification of the color field of a cluster of overlapped disks.

\begin{figure}
\centering
\includegraphics[scale=0.5]{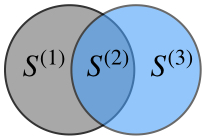}
\caption{Scheme of the area covered by two overlapped strings.}
\label{fig:ov2}
\end{figure}

\subsection{Color field of overlapped strings and the color suppression factor}
We start this diskussion assuming that the emission of $q\bar{q}$ pairs from the cluster of strings proceeds independently and it is governed by the strength of the color field of the cluster.
Evidently, this cluster has not only different color field than the individual strings that shape it but also different transverse area.
First, it is necessary to analyze the case of an isolated string. In this case, a single string with transverse area $S_1$ emits partons with transverse momentum distribution
\begin{equation}
\frac{d\sigma}{dyd^2p}=C\exp\left( -\frac{m_T^2(p_T)}{t_1} \right),
\end{equation}
where $t_1$ is the tension of the string and $m_T^2=m^2+p_T^2$, being $p_T$ and $m$ the transverse momentum and mass of the emitted parton. The tension, according to the Schwinger mechanism, is proportional to the field, and thus to the color charge \cite{BIRO1984449,BIALAS1986242}, which we denote by $Q_0$. The mean square transverse momentum is $\langle p_T^2\rangle_1=t_1$ and proportional to $Q_0$.
Then the density of color charge is $q=Q_0/S_1$.
We denote the mean multiplicity of produced particles by a single string per unit of rapidity as $\mu_1$ which is also proportional to the color charge.
Now, we compute the case of two overlapped strings.
In this scenary, it is expected that the total production of charged particles comes from the three independent regions $S^{(1)}$, $S^{(2)}$ (overlapping region) and $S^{(3)}$ depicted in Fig.~\ref{fig:ov2}.
Note that the regions are not independent and can be related as follows $S^{(3)}=S^{(1)}=S_1-S^{(2)}$.
Then, the color in each of the non overlapping areas will be
$Q_1=Q_0S^{(1)}/S_1$.
On the other hand, the total color in the overlap area $Q_2$ will be a vector sum of the two overlapping colors $qS^{(2)}$.
In this summation the total color squared should be conserved \cite{PhysRevLett.85.4864}.
Thus $Q_2^2=(\mathbf{Q}_{ov}+\mathbf{Q}_{ov}')^2$, where $\mathbf{Q}_{ov}$ and $\mathbf{Q}_{ov}'$ are the two vector colors in the overlap area. Since the colors in the two strings may generally be oriented in arbitrary directions respective and independent to one another, the average $\langle \mathbf{Q}_{ov}\mathbf{Q}_{ov}' \rangle$ is zero, which leads to $Q_2=\sqrt{2} q S^{(2)}=\sqrt{2} Q_0 S^{(2)}/S_1$ \cite{Braun2000}.
Notice that due to the vector nature, the color in the overlap is less than the sum of the two overlapping colors.
This effect has important consequences concerning the saturation of multiplicities and the rise of the mean transverse momentum with multiplicity.
Thus, assuming independent emission from the three regions of Fig.~\ref{fig:ov2}, we obtain for the multiplicity weighted by the multiplicity for a single string ($\mu_1$)
\begin{equation}
\mu/\mu_1=2(S^{(1)}/S_1)+\sqrt{2}(S^{(2)}/S_1),
\end{equation}
and for the mean square transverse momentum (we divide the total transverse momentum by the multiplicity)
\begin{eqnarray}
\frac{\langle p_T^2 \rangle}{\langle p_T^2 \rangle_1}&=&\frac{2(S^{(1)}/S_1)+\sqrt{2}\sqrt{2}(S^{(2)}/S_1)}{2(S^{(1)}/S_1)+\sqrt{2}(S^{(2)}/S_1)} \nonumber\\
 &=&\frac{2}{2(S^{(1)}/S_1)+\sqrt{2}(S^{(2)}/S_1)},
\end{eqnarray}
where we have used the property $S^{(1)}+S^{(2)}=S_1$. Generalizing to any number $N$ of overlapping strings, we have
\begin{eqnarray}
\frac{\mu}{\mu_1}&=&\sum_i \sqrt{n_i}(S^{(i)}/S_1),\\
\frac{\langle p_T^2 \rangle}{\langle p_T^2 \rangle_1} &=&\frac{\sum_i (S^{(i)}/S_1)}{\sum_i \sqrt{n_i}(S^{(i)}/S_1)}=\frac{N}{\sum_i \sqrt{n_i}(S^{(i)}/S_1)},\label{eq30}
\end{eqnarray}
where the sum runs over all individual overlaps of $n_i$ strings having areas $S^{(i)}$.
We have used the identity $\sum_iS^{(i)}=NS_1$. These equations are not easy to apply because we have to identify all individuals overlaps of any number of strings with their areas.
However one can avoid these difficulties realizing that one can combine all terms with a given number of overlapping strings $n_i=n$ into a single term, which sums all such overlaps into a total area of exactly $n$ overlapping strings $S_n^{Tot}$.
Then, one can write
\begin{eqnarray}
\frac{\mu}{\mu_1}&=&\sum_{n=1}^N \sqrt{n}(S_n^{Tot}/S_1),\\
\frac{\langle p_T^2 \rangle}{\langle p_T^2 \rangle_1} &=&\frac{N}{\sum_{n=1}^N \sqrt{n}(S_n^{Tot}/S_1)}.
\end{eqnarray}
The total area can be determined in the thermodynamic limit. One finds that the distribution of overlap strings over the total surface $S$ in the variable $n$ is Poissonian with mean $\rho=NS_1/S$, which corresponds to the filling factor in the percolation context. Therefore, the fraction of the total area covered by strings will be $1-\exp(-\rho)$, which matches with the corresponding fraction of area covered by disks in two-dimensional continuum percolation in Eq.~\eqref{eq:CAD-TL}. Note that the multiplicity in Eq.~\eqref{eq30} is damped by a factor
\begin{equation}
F_\text{TL}(\rho)=\sqrt{\frac{\phi_\text{TL}(\rho)}{\rho}},
\label{eq-EF}
\end{equation}
which can be interpreted as a factor that suppress the total color charge of overlapped strings. In the following, we name $F(\rho)$ as the color suppression factor.
Finally, we can write for the mean values of the multiplicity and the square transverse momentum as \cite{Braun2015,RMF,Braun2000}
\begin{eqnarray}
\mu & = & NF_\text{TL}(\rho)\mu_1, \label{eq38}\\
\langle p_T^2 \rangle & = & \langle p_T^2 \rangle_1/F_\text{TL}(\rho) \label{eq39}.
\end{eqnarray}
Note that both the multiplicity and the transverse momentum distributions depend explicitly on the color suppression factor, and therefore in the fraction of area covered by disks. Then any modification from any source in the latter leads corrections in the phenomenological description of the experimental results in high energy physics.

\subsection{Modifications to the color suppression factor}

\begin{figure*}
\centering
\includegraphics[scale=0.2]{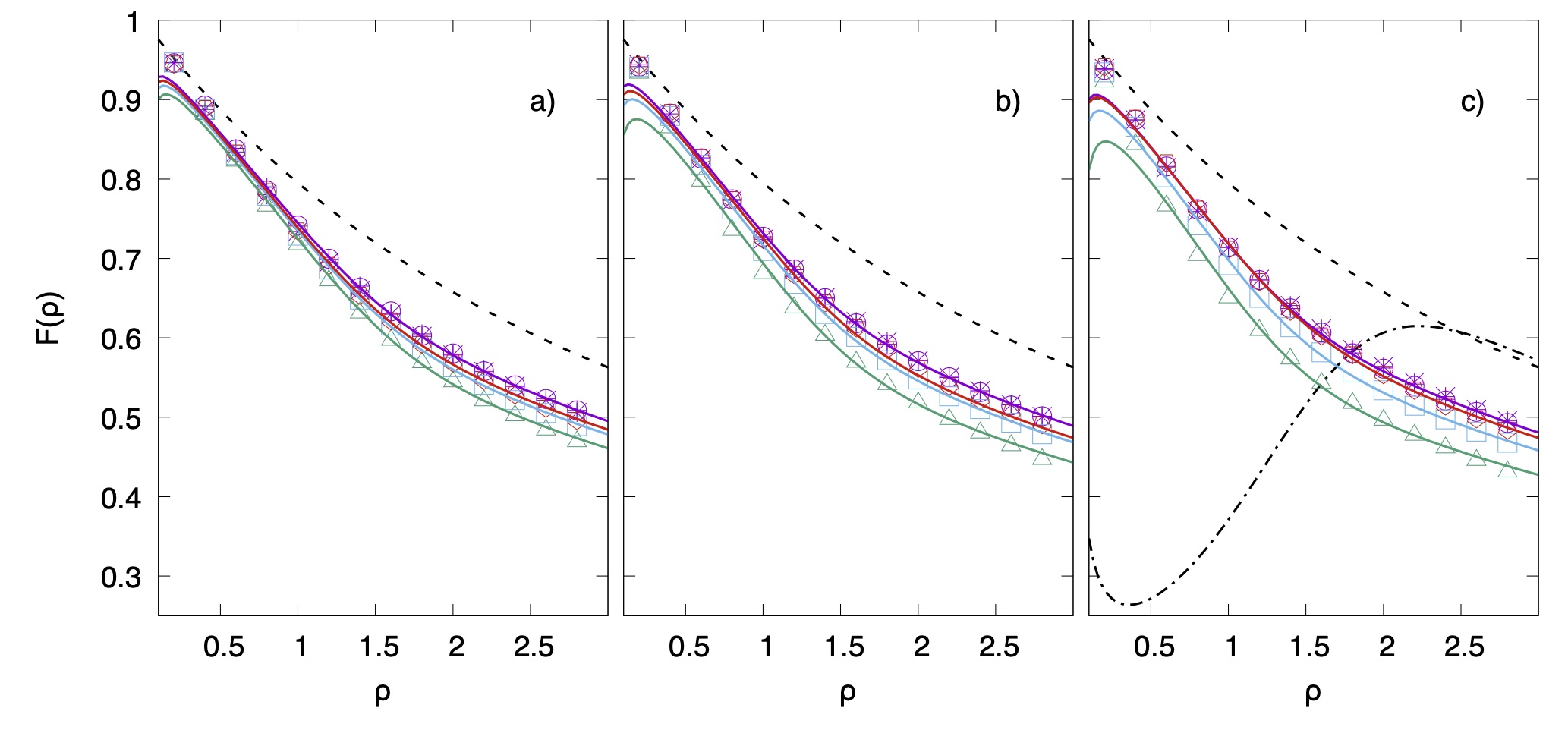}
\caption{Comparison between the color suppression factor computed from the simulation data (figures) and the computed from the fit function (solid lines) on Eq.~\eqref{eq:df} for small-bounded string systems for $N=$13, and bounded by circles (circles), ellipses (crosses and stars), triangles (triangles), squares (squares) and pentagons (pentagons), and for the a) U, b) 1S and c) SQ density profiles. Dashed lines and dash-dot line are the corresponding color suppression factor in the thermodynamic limit for the uniform and SQ density profiles.}
\label{fig:Fes}
\end{figure*}

\begin{figure*}
\centering
\includegraphics[scale=0.2]{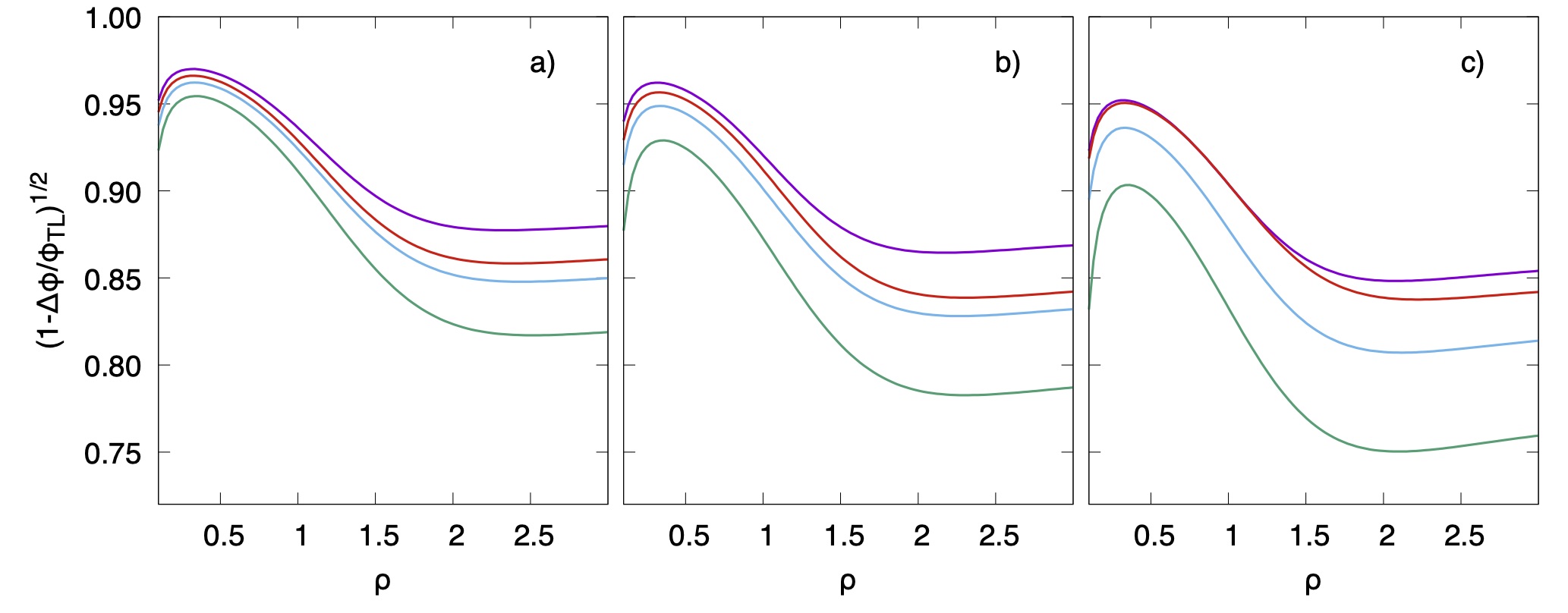}
\caption{Damping factor due to small-bounded effects in string systems for $N=13$ all shapes of the boundary e2 (purple line), e3 (green line), e4 (blue line), and e5 (red line), for the density profiles: a) U, b) 1S, and c) SQ.}
\label{fig:Ft}
\end{figure*}

As we diskussed in Sec. \ref{results}, the fraction of area covered by disks deviates from $\phi_\text{TL}$ due to the size and the boundary shape of the percolating system.
Assuming the same arguments to deduce the color suppression factor, we propose change $\phi \to \phi_\text{TL}$ in Eq.~\eqref{eq-EF}, thus
\begin{equation}
    F(\rho)=\sqrt{\frac{\phi(\rho)}{\rho}},
\end{equation}
is the color suppression factor for small-bounded systems.
The latter can be expressed in terms of $\Delta \phi$, $\phi_\text{TL}$ and $F_\text{TL}$ as follows
\begin{equation}
    F(\rho)=F_\text{TL}(\rho)\left(1-\frac{\Delta \phi(\rho)}{\phi_\text{TL}(\rho)}  \right)^{1/2},
\label{eq:Ffact}
\end{equation}
where the second factor in the right-hand side may be interpreted as a damping function of the color field due to small-bounded effects, which becomes relevant for small percolating systems.
For example, in the case of $N=$13, where the maximum deviation of the area covered is observed, the data of the color suppression factor is well fitted by the expression in \eqref{eq:Ffact} for $0.4<\rho<3$, as we shown in Fig.~\ref{fig:Fes}. 
It is remarkable since the most interesting phenomena in high energy physics are observed in systems with densities in the range $0.5<\rho<2$ for the pp case.
We also show in Fig.~\ref{fig:Ft} the explicit dependence of this damping function on the filling factor for all shapes of the boundary and all density profiles for string system with number of disks $N=13$.
The value $N=$13 is close to the values found for minimum bias in pp collisions at $\sqrt{s}=$7TeV.
The deviation of the color suppression factor from its value in the thermodynamic limit is given by
\begin{equation}
    \frac{F_\text{TL}(\rho)-F(\rho)}{F_\text{TL}(\rho)}=1-\left(1-\frac{\Delta \phi(\rho)}{\phi_\text{TL}(\rho)}  \right)^{1/2}.
\end{equation}
In both Fig.~\ref{fig:Fes} and Fig.~\ref{fig:Ft}, we have obviated the error bars from the propagated error since the relative error due to the fit parameter in $\Delta \phi$ are of the order of 0.3\%, which are technically unperceived.

Due to the form of $\Delta \phi$, the relative deviation starts at a minimum value. Then, for $\rho^*-L/2<\rho<\rho^*+L/2$, it increases linearly until reaching a maximum value, where it will hold constantly for $\rho^*+L/2<\rho<3$. 
In the region of high densities, the percentage of the deviation takes values around 11\%-19\% for string systems with disks uniformly distributed. However, in the case of the Gaussian density profile, where the disks are more concentrated in the geometrical center of the boundary, this percentage becomes higher, taking values around 15\%-25\%, for ellipses and triangular boundaries, respectively.

\begin{figure*}
\centering
\includegraphics[scale=0.2]{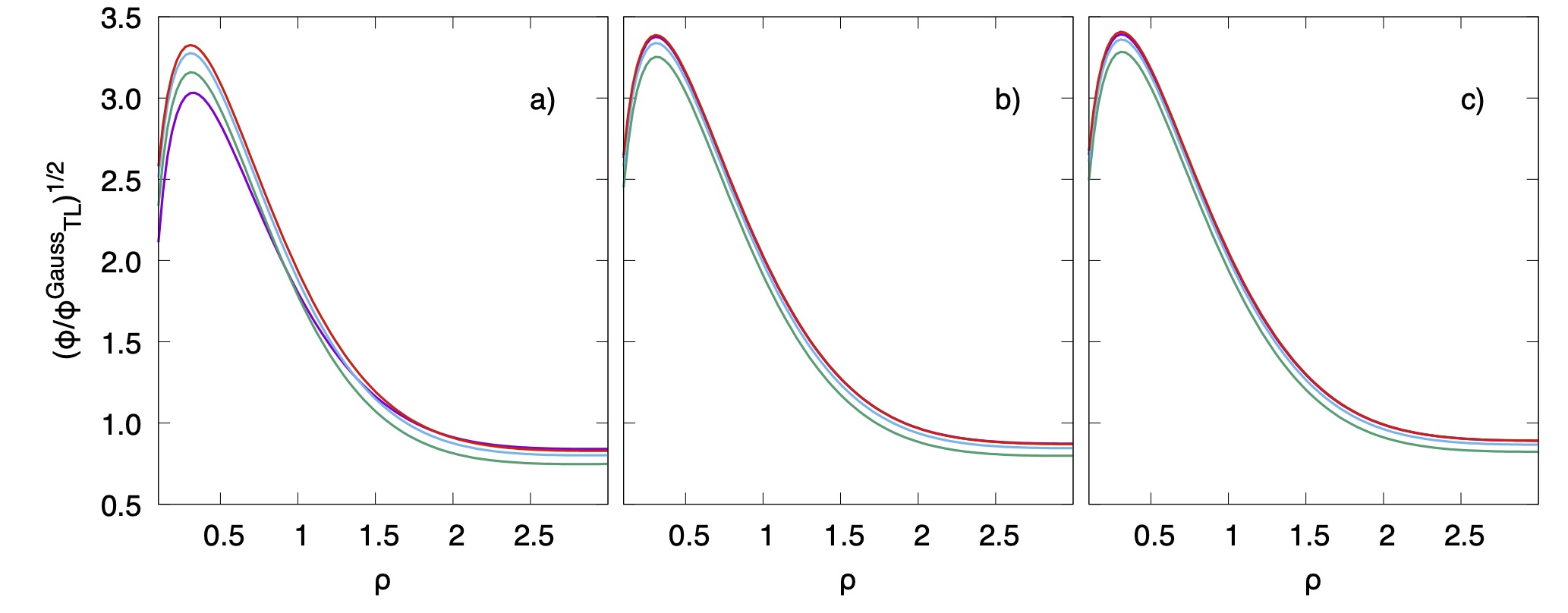}
\caption{Damping factor due to small-bounded string systems for the color suppression factor in the SQ density profile for a) $N=$13, b) $N=$55 y c) $N=$96 bounded by circles and ellipses (purple line), triangles (red line), squares (cyan line) and pentagons (green line).}
\label{fig:Ftgauss}
\end{figure*}

On the other hand, it is possible to determine the corresponding deviation of the fraction of area covered by disks in the Gaussian density profile from its value in the thermodynamic limit for the same Gaussian profile in Eq.~\eqref{eq:CAD-sq} using the data for $a^\text{Gauss}$, $b^\text{Gauss}$ and $\rho_c$ reported in the literature for pp and pA collisions. 
In this case, we analyze only the systems constituted by $N$=13, 55 and 96 strings. 
In Fig.~\ref{fig:f} c) we plot the thermodynamical limit for both uniform (dashed line) and Gaussian (dash-dot line) density profiles together the simulation data.

Similarly as we have determined the deviation of the fraction of area covered by disks for $\phi$ from $\phi_\text{TL}$, we define
\begin{equation}
    \Delta \phi ^\text{Gauss}(\rho)=\phi^\text{Gauss}_\text{TL}(\rho)-\phi(\rho),
\end{equation}
where $\phi$ is the fraction of area covered by disks measured by simulation and $\phi^\text{Gauss}_\text{TL}$ is its corresponding value in the thermodynamic limit for the Gaussian profile in Eq.~\eqref{eq:CAD-sq}.
As $\phi$ is well characterized by Eq.~\eqref{eq:f-cor} as a function of $\phi_\text{TL}$ and $\Delta \phi$, and its corresponding fit parameters are also determined (see Table~\ref{tab:pfits}), then $\Delta \phi^\text{Gauss}$ is well determined and it fits correctly the simulation data.
In this way, we re-write the color suppression factor as follow
\begin{equation}
    F(\rho)=F^\text{Gauss}_\text{TL}(\rho)\left(1-\frac{\Delta \phi^\text{Gauss}(\rho)}{\phi^\text{Gauss}_\text{TL}(\rho)}  \right)^{1/2},
\label{eq:Ffact-gauss}
\end{equation}
where $F^\text{Gauss}_\text{TL}$, as usual in the string percolation model, is computed by replacing $\phi_\text{TL}^\text{Gauss}$ instead $\phi_\text{TL}$ in Eq.~\eqref{eq-EF}.
This last expression has the same structure than the color suppression factor in  Eq.~\eqref{eq:Ffact}, where it has been replaced the functions $F_\text{TL}\to F^\text{Gauss}_\text{TL}$, $\Delta \phi \to \Delta \phi^\text{Gauss}$ and $\phi_\text{TL}\to \phi^\text{Gauss}_\text{TL}$ defined for the Gaussian density profile. 
In the same way as in Eq.~\eqref{eq:Ffact}, the second factor in the right-hand side can be interpreted as a second damping factor due to small-bounded effects.
Substituting the explicit form of $\Delta \phi ^\text{Gauss}_\text{TL}$, we found that the damping factor is re-written as $(\phi/\phi^\text{Gauss}_{TL})^{1/2}$, which is plotted in Fig.~\ref{fig:Ftgauss}.
In this way, the deviation of $F$ from $F^\text{Gauss}_{TL}$ is given by
\begin{equation}
    \frac{F^\text{Gauss}_\text{TL}(\rho)-F(\rho)}{F^\text{Gauss}_\text{TL}(\rho)}=1-\left(  \frac{\phi(\rho)}{\phi^\text{Gauss}_\text{TL}(\rho)}\right)^{1/2},
\end{equation}
which is well defined through known functions.
Notice that in the case of low-density systems ($\rho<1$), the relative deviation is large, taking values between 100\%-150\%. However, for $1<\rho<2$, this deviation becomes smaller as a crossover effect of $F^\text{Gauss}_\text{TL}$ and the simulation data. Finally, for high density percolating systems, the deviation reaches a plateau region, and it becomes very similar to the case of the uniform density profile since for high-density systems there are not substantial differences between $F_\text{TL}$ and $F^\text{Gauss}_\text{TL}$, as can be seen in Fig.~\ref{fig:Fes} c).

\subsection{Color suppression factor at $\rho_\text{max}$}

As we mentioned in Sec. \ref{results}, the filling factor is physically upper bounded in small-bounded percolating systems and the corresponding fraction of the area covered is given by the packing fraction. Then, substituting in Eq.~\eqref{eq-EF}, we found the color suppression factor at the maximum filling factor is
\begin{equation}
    F(\rho_\text{max})=\frac{1}{\sqrt{N}}.
\end{equation}
Note that the color suppression factor at $\rho_\text{max}$ is a function depending only on the number of strings in the system.

\subsection{Temperature in string percolating systems}

It is well-known that the distribution of transverse momentum of the charged particle produced in high energy physics experiments can be determined by first principles.
According to the Schwinger mechanism for the particle production, the distribution for massless particles expressed in terms of $p_T^2$ and the string tension $x$ is given by
\begin{equation}
\frac{dn}{dp_T^2}\sim \exp\left(-\pi\frac{p_T^2}{x^2}  \right).
\label{eq:dS}
\end{equation}

The superposition of clusters with different number of strings each satisfying Eq.~\eqref{eq:dS} gives rise to the transverse momentum distribution (the cluster size distribution is a gamma function)
\begin{equation}
\frac{dn}{dp_T^2}\sim \left(1+\frac{p_T^2}{k\langle p_T^2\rangle} \right)^{-k}=
\left(1+\frac{F_\text{TL}(\rho)p_T^2}{k\langle p_T^2\rangle_1} \right)^{-k}
\label{eq:ptTsallis}
\end{equation}
which describes rightly the pp experimental data. At low $p_T$, the Eq.~\ref{eq:ptTsallis} becomes
\begin{equation}
\frac{dn}{dp_T^2}\sim \exp \left(-\frac{F_\text{TL}(\rho)p_T^2}{\langle p_T^2\rangle_1 }\right).
\label{eq:thermal}
\end{equation}
On the other hand, inside each cluster, the chromoelectric field is not constant and then the tension fluctuates around its mean value.
This fluctuation is related to the stochastic picture of the QCD vacuum. Since the average value of the color field strength must vanish, it can not be constant but change randomly from point to point. Such fluctuations lead to a Gaussian distribution of the string tension \cite{BIALAS1999301}. Then, as
\begin{eqnarray}
\frac{dn}{dp_T^2}&\sim& \sqrt{\frac{2}{\langle x^2 \rangle}} \int_0^\infty dx \exp \left( -\frac{x^2}{2 \langle x^2 \rangle}  \right) \exp\left(-\pi \frac{p_T^2}{x^2}\right)\nonumber\\ 
&\sim& \exp\left(-p_T\sqrt{\frac{2\pi}{\langle x^2 \rangle}} \right),\label{eq:intGauss}
\end{eqnarray}
we obtain a thermal distribution. Using Eq.~\eqref{eq:intGauss}, the distribution in Eq.~\eqref{eq:thermal} becomes
\begin{equation}
\frac{dn}{dp_T^2}\sim \exp\left(-\frac{p_T}{T_\text{TL}(\rho)}\right),
\end{equation}
with
\begin{equation}
T_\text{TL}(\rho)=\sqrt{\frac{\langle p_T^2  \rangle_1}{2F_\text{TL}(\rho)}}.
\label{eq:T}
\end{equation}
Therefore in the string percolation framework, the fluctuations of the string tension give rise to the thermal distribution of the transverse momentum with a temperature which can be interpreted as the temperature of the initial state of the system \cite{Scharenberg2011}.

If we take as the phase transition temperature the experimentally determined chemical freeze-out temperature $T_c=167.7\pm2.6$ MeV for the percolation threshold $\rho_c=1.2$, $F_\text{TL}(\rho_c)=0.66$ and using Eq.~\eqref{eq:T} we obtain $\sqrt{\langle p_T^2  \rangle_1}=207.2\pm3.3$MeV, which is close to the value 200MeV used in the phenomenological applications \cite{Scharenberg2011}.

Taking into account that the color suppression factor is damped for small-bounded systems, this temperature is also corrected by the damping factor $(1-\Delta \phi/ \phi_\text{TL})^{-1/4}$.
Moreover, at $\rho_\text{max}$, the behavior of the temperature as a function of the number of string goes as $T({\rho_\text{max}})\propto N^{1/4}$, and becomes independent of the geometry and the density profile.
Note that Eq.~\eqref{eq:T} allows us to define a critical temperature $T_\text{c}$ associated with the percolation threshold as $T_\text{c}=T(\rho_\text{c})$.
In Sec.~\ref{SS}, we use this temperature definition to analyze the small-bounded effects in the speed of sound (in the high energy physics context) through the modifications to the color suppression factor considering the percolation threshold for small and elliptically bounded systems reported in Ref.~\cite{jerc}.

\subsection{Critical behavior of the speed of sound}\label{SS}

\begin{figure}
\centering
\includegraphics[scale=0.2]{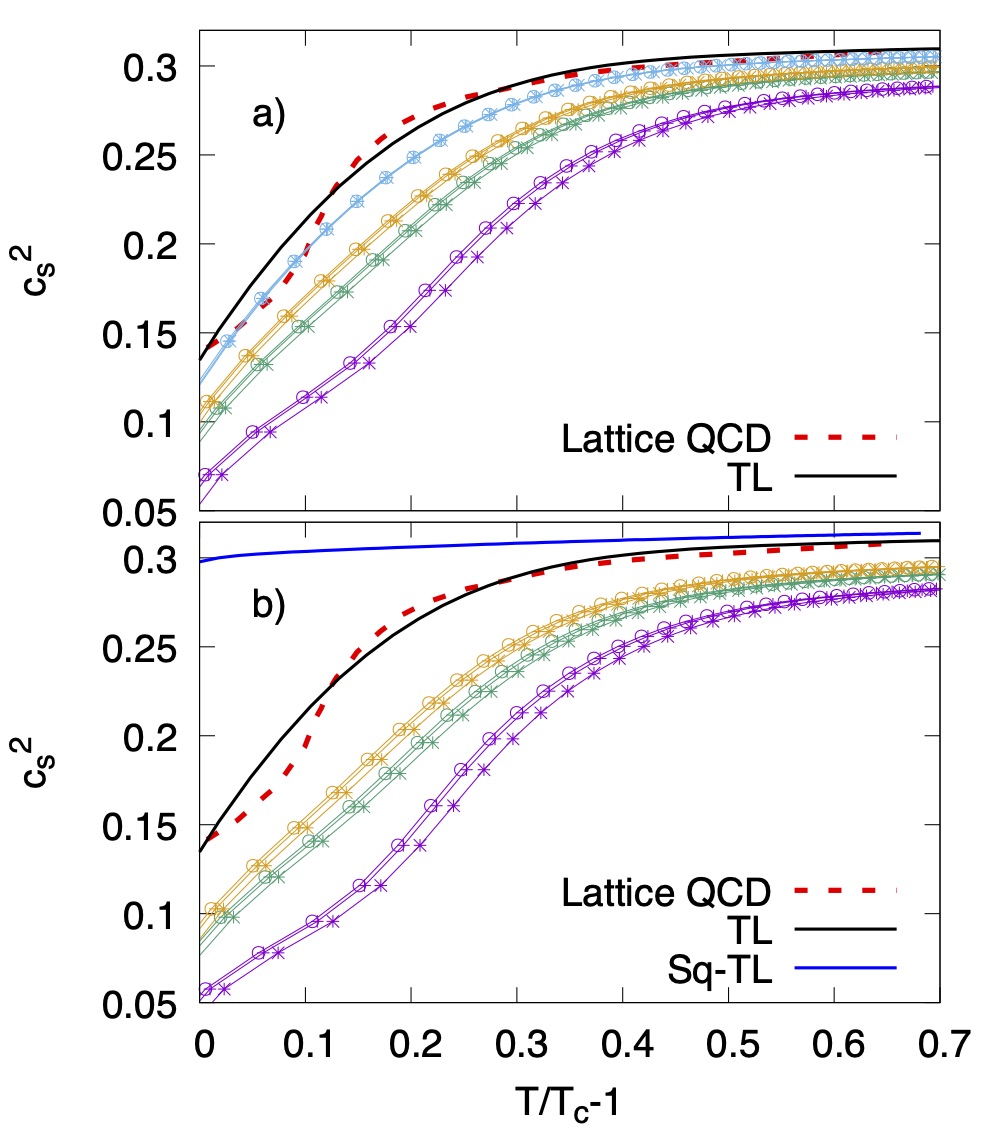}
\caption{Speed of the sound for small string systems bounded by ellipses for $N=$13 (purple), 55 (green), 96 (yellow) and 500 (cyan), for different density profiles: a) U and b) SQ. Solid lines correspond to the speed of sound in the thermodynamic limit for the U (black) and Sq (blue) density profiles. Red dashed line is the lattice QCD prediction.}
\label{fig:cs}
\end{figure}

The most important feature of the string percolation model is its ability to predict several observables in the high energy physics for both theoretical and experimental results. A completed review on the string percolation applications can be found in Refs.~\cite{Braun2015, RMF}. Particularly, coupling this framework with the 1D Bjorken expansion, it is possible to compute the speed of sound as \cite{Scharenberg2011,Braun2015}
\begin{equation}
c_s^2=\frac{1}{3}\left(\frac{\exp(-\rho)}{F_\text{TL}^2}-1 \right) \left( 0.0191 \frac{1}{\rho F_\text{TL}^2} \left( \frac{\eta}{s} \right)_\text{TL}^{-1}-1 \right),
\end{equation}
where $\eta$ is the shear viscosity and $s$ is the entropy density, which in the context of string percolation (in the thermodynamic limit) is given by 
\begin{equation}
\left(\frac{\eta}{s} \right)_\text{TL}=\frac{T_\text{TL}L}{5\phi_\text{TL}},
\end{equation}
where $L=1$fm is the longitudinal extension of the string. 
In the following, we replace $F_\text{TL}\to F$ and $\phi_\text{TL}\to\phi$ in order to include the modifications due to small-bounded effects in the speed of sound.
We have obviated the explicit dependence on $\rho$ in order to improve notation. 
We also define the order parameter
\begin{equation}
\tau=\frac{T}{T_\text{c}}-1,
\label{eq:tau}
\end{equation}
for $T>T_\text{c}$, where $T_\text{c}$ is the critical temperature, which is computed as the evaluation of $T$ in the percolation threshold. 
As far as we know, the percolation threshold has been reported for both U and SQ density profiles in the thermodynamic limit (see Refs.~\cite{Quintanilla_2000,PhysRevE.76.051115,PREMertens,RODRIGUES1999402}) and small and elliptical bounded systems considering several density profiles in Ref.~\cite{jerc}.
In Fig.~\ref{fig:cs}, we show the behavior of the speed of sound as a function on the number of disks for the Uniform (Fig.~\ref{fig:cs} a)) and SQ (Fig.~\ref{fig:cs} b)) and elliptically bounded string systems considering the eccentricities $\varepsilon=$0.0, 0.5 and 0.9 together its corresponding function in the thermodynamic limit and the lattice QCD prediction.
For $N$ fixed, we observe a smooth dependence of $c_s^2$ on the eccentricity even though the fraction of area covered by disks does not show differences in any cases of ellipses analyzed. This is due to the percolation threshold shows an explicit dependence on the eccentricity for small systems ~\cite{jerc}.

\begin{figure}
\centering
\includegraphics[scale=0.2]{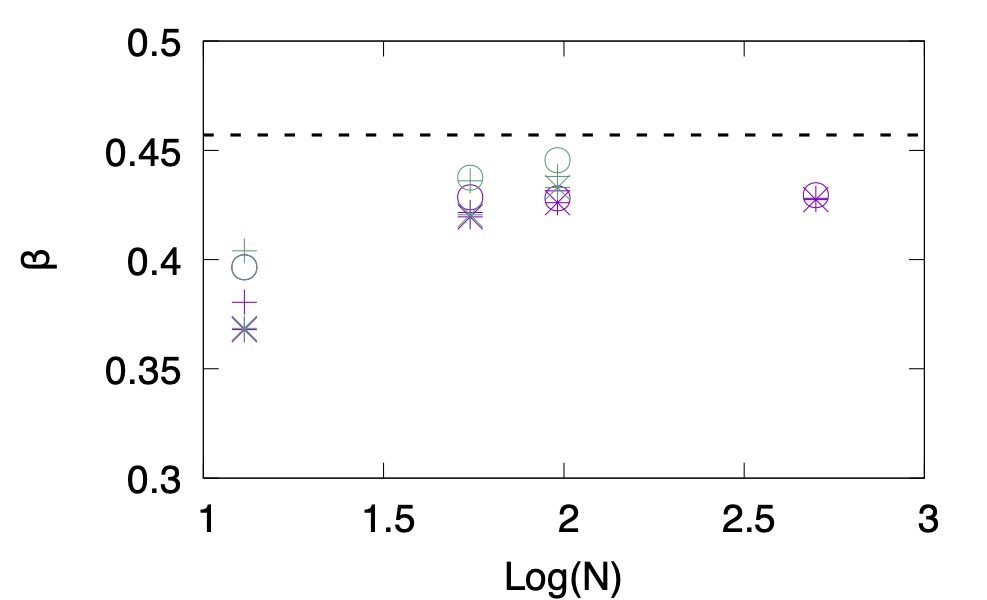}
\caption{Critical exponent of the power laws behavior of the speed of sound around the critical temperature for small and elliptically bounded string systems as a function of $N$ and the eccentricity: $\varepsilon=$0.0 (circles), 0.5 (crosses) and 0.9 (stars), for the Uniform (purple figures) and SQ (green figures) density profiles.}
\label{fig:esc-cs}
\end{figure}

Moreover, it is possible to determine the critical behavior of the speed of sound around the critical temperature. 
This is done by fitting the data of $c_s^2$ in all curves in Fig.~\ref{fig:cs} for $0<\tau<0.1$. 
We observe that $c_s^2$ follows a power law scaling for $T>T_\text{c}$ given by
\begin{equation}
c_s^2\sim \tau^{2\beta}.
\end{equation}
Figure~\ref{fig:esc-cs} shows the values of $\beta$ as a function of the number of disks for elliptically bounded systems for both uniform and Gaussian SQ density profiles. 
Note that this exponent grows as $N$ increases. This last effect may be understood as the capability of the string system to add more disks to the spanning cluster and becomes denser.
Particularly, we found $\beta=0.456\pm0.003$ in the thermodynamic limit for the uniform density profile, which is close to the value 1/2 predicted by the mean field theory for magnetization around the critical temperature in the Ising model. 

On the other hand, it is possible to determine the speed of sound for high density string systems. At $\rho_\text{max}$, we found
\begin{eqnarray}
\rho_\text{max}\sim N ,& F(\rho_{max}) \sim N^{-1/2}, \nonumber\\
T(\rho_\text{max}) \sim N^{1/4},& (\eta/s)(\rho_\text{max})\sim N^{1/4}.
\end{eqnarray}
Therefore, the speed of sound at $\rho_\text{max}$ can be written as a function that only depends on the number of disks:
\begin{equation}
c_s^2(\rho_\text{max})=\frac{1}{3}\left(N\exp(-N)-1 \right)(\text{const} N^{-1/4}-1).
\end{equation}
Taking the limit $N\to \infty$ we found that $c_s^2\to 1/3$, which is in a full agreement with the reported in the literature \cite{BOYD1996419,PhysRevLett.101.181601}.

\section{Conclusions}\label{Conclusions}

In this work, we have presented the algorithm to build a 2-dimensional continuum percolating system free of periodic boundary conditions and delimited by different geometries, with a low population of disks and filled with mainly two different density profiles: Uniform and Gaussian. The second one is relevant in the string percolation since the nuclear profile is denser in the center of the overlapping area than the region near to the border. We also provided the Monte Carlo integration method to measure the fraction of area covered by disks in such percolating systems.

Simulation results indicate an explicit dependence of the fraction of area covered by disks on the number of disks, boundary shape, and the density profile, which is more relevant in small systems.
However, the percolating system lost this dependence as $N$ increases. This is beacause there is no difference between very large circles (compared with the disk size) and very large ellipses or very large regular polygons and thus the system behaves as in the thermodynamic limit.
Other limit analyzed corresponds to the situation when all disks in the percolating system are concentric with the geometric shape of the boundary. In this limit, the percolating system reaches its maximum filling factor value and the fraction of area covered by disks is exactly the packing fraction, becoming independent of the number of disks and the density profile.

The modifications on the fraction of area covered by disks in small and bounded systems lead corrections to the color suppression factor in the string percolation model as a second damping function of the color due to finite size and boundary effects.
This function is well determined by Eqs.~\eqref{eq:Ffact} and \eqref{eq:Ffact-gauss}.
In the case of smallest system analyzed ($N=$13), this damping involves relatives errors of the order between 11\% and 25\% (depending on the filling factor and boundary shape) for the uniform density profile and high-density systems (1.5$<\rho$). On the other hand, for low-density systems ($\rho<$0.5), we can assure that the area covered by disks is well determined by its value in the thermodynamic limit taking into account that it includes a relative error of the order of 5\%. However, these results change drastically for the Gaussian density profile. Particularly, for low-density systems ($\rho<$0.75), the damping factor takes values greater than 2.5, it means that the absolute relative error is of the order between 150\% and 250\%.
In spite of that, the second damping factor becomes more similar to the corresponding to the uniform density profile since the color suppression factor in the thermodynamic limit for both uniform and Gaussian profiles look very similar in high-density systems ($\rho>2.3$). Moreover, at $\rho_\text{max}$, the color suppression factor just only depends on the number of strings distributed in the overlapping surface.

Although the color suppression factor has low deviations from its value in the thermodynamic limit even in the case of systems populated by 13 strings, it is necessary to take into account the finite size-shape effects of all quantities involved to compute any observable, which could give rise to significant deviations from the determination in the thermodynamic limit.
For example, the speed of sound shows a dependence on both the strings number and on the eccentricity around the critical temperature for elliptically bounded systems even though the fraction of area covered by disks does not have notable deviations as a function of the eccentricity. This is due to the dependence of the percolation threshold on the number of disks and the eccentricity, which carry on modifications of the critical temperature.
 Then, in the temperature scale, the speed of sound is shifted as a function on the number of disks and the boundary shape.
 On the other hand, we compute the speed of sound at $\rho_\text{max}$, which just only depends on the number of disks  as is expected. Moreover, in the limit of high populated systems ($N\to\infty$), the speed of sound takes the value 1/3, which is in a full agreement with the reported by the calculations of lattice QCD.

Additionally, we determine the behavior of the speed of sound around the critical temperature in the thermodynamic limit, finding that the corresponding critical exponent of the power law is 0.45, which is close to the value 1/2 for the critical exponent $\beta$ of the magnetization in the Ising model determined by the mean field theory. This is an indication that the speed of sound is related to the capability of the spanning cluster to add more disks or finite clusters as the temperature increase.
On the contrary, for the Gaussian density profile, this exponent takes the value 0.2, which means that the spanning cluster is constituted by a large proportion of disks in the system, then, the speed of sound reaches values close to limit value 1/3 as a consequence of the saturation of the spanning cluster.
 These behaviors of both the uniform and Gaussian density profiles could be verified by analyzing the percolation susceptibility as a function of the temperature around the percolation threshold.
In the same sense, it is worth to mention that the filling factor and the temperature for string systems define different scales. Therefore there is not a direct way to determine the correlation length for both parameters, and it is not possible to establish a scaling hypothesis using the critical exponents determined for two-dimensional percolation. To do this, it is necessary to determine the corresponding exponent $\nu$ associated with the correlation length, considering the parameter $\tau$ of Eq.~\eqref{eq:tau} as the order parameter.

At RHIC and LHC energies the string density for minimum bias pp collisions is in the range $0.3<\rho<0.6$ and the differences for the suppression factor between the thermodynamic limit value and the corresponding profile already taking into account the finite-shape effects is less than 5\% (see Fig.~\ref{fig:Fes}), the effects concerning global observables like multiplicity or transverse momentum distribution is negligible. Even in the case of high multiplicity events where the filling factor takes values larger than 1 or 1.5, the difference are of the order of 10\% which are comparable to the uncertainties of the comparison with experimental data. However, the differences can be experimentally seen looking at harmonics of the azimuthal distribution and some correlations between them.
In Fig.~\ref{fig:Ft} we show that at high multiplicity, $\rho>1.5$, the difference in the eccentricity $e_3$ are of the order of 25\% in the cases of an extended gaussian profile.

\section{Acknowledgments}
J.E.R. acknowledges financial support from CONACYT (postdoctoral fellowship Grant no. 289198).
C.P. was supported by the grant Maria de Maeztu Unit of Excellence MDM-20-0692 and FPA project 2017-83814-P of Ministerio de Ciencia e Inovación of Spain, FEDER and Xunta de Galicia. 


\end{document}

%% file: tabla.tex
\begin{table*}[!ht]
\caption{Fit parameters of $\Delta \phi$ (Eq.~\eqref{eq:df}).}\label{tab:pfits}

\begin{center}
\resizebox{\textwidth}{!}{
\begin{tabular}{c c |c c c| c c c| c c c}

\hline
 & & &$a$ & & & $\rho^*$ & & &$\Delta L$ & \\\cline{3-5}\cline{6-8}\cline{9-11}
 $N$ & Geom. & U & 1S & SQ & U & 1S & SQ & U & 1S & SQ\\
\hline
13&	e2&	0.108	$\pm$0.002&	0.117	$\pm$0.002&	0.129	$\pm$0.002&	1.20	$\pm$0.02&	1.12	$\pm$0.02&	1.07	$\pm$0.02&	0.70	$\pm$0.04&	0.68	$\pm$0.03&	0.68	$\pm$0.04\\
 &	e3&	0.158	$\pm$0.003&	0.182	$\pm$0.004&	0.202	$\pm$0.004&	1.25	$\pm$0.03&	1.13	$\pm$0.03&	1.03	$\pm$0.03&	0.75	$\pm$0.05&	0.75	$\pm$0.05&	0.73	$\pm$0.05\\
 &	e4&	0.133	$\pm$0.002&	0.147	$\pm$0.003&	0.161	$\pm$0.003&	1.23	$\pm$0.03&	1.14	$\pm$0.03&	1.07	$\pm$0.02&	0.73	$\pm$0.04&	0.72	$\pm$0.05&	0.70	$\pm$0.04\\
 &	e5&	0.124	$\pm$0.002&	0.131	$\pm$0.002&	0.139	$\pm$0.003&	1.22	$\pm$0.02&	1.17	$\pm$0.02&	1.12	$\pm$0.03&	0.71	$\pm$0.03&	0.71	$\pm$0.04&	0.71	$\pm$0.05\\

\hline

55&	e2&	0.079	$\pm$0.002&	0.091	$\pm$0.002&	0.104	$\pm$0.002&	1.42	$\pm$0.03&	1.37	$\pm$0.03&	1.30	$\pm$0.03&	0.87	$\pm$0.05&	0.87	$\pm$0.05&	0.85	$\pm$0.05\\
 &	e3&	0.103	$\pm$0.003&	0.135	$\pm$0.003&	0.165	$\pm$0.004&	1.43	$\pm$0.04&	1.30	$\pm$0.03&	1.18	$\pm$0.03&	0.86	$\pm$0.05&	0.84	$\pm$0.05&	0.83	$\pm$0.06\\
 &	e4&	0.092	$\pm$0.003&	0.108	$\pm$0.002&	0.127	$\pm$0.003&	1.45	$\pm$0.04&	1.36	$\pm$0.03&	1.25	$\pm$0.03&	0.88	$\pm$0.06&	0.84	$\pm$0.05&	0.83	$\pm$0.05\\
 &	e5&	0.086	$\pm$0.002&	0.096	$\pm$0.002&	0.106	$\pm$0.003&	1.44	$\pm$0.04&	1.40	$\pm$0.03&	1.34	$\pm$0.04&	0.86	$\pm$0.05&	0.84	$\pm$0.05&	0.84	$\pm$0.05\\
 
 \hline
 
 96&	e2&	0.066	$\pm$0.002&	0.075	$\pm$0.002&	0.088	$\pm$0.002&	1.50	$\pm$0.04&	1.42	$\pm$0.03&	1.33	$\pm$0.03&	0.94	$\pm$0.05&	0.90	$\pm$0.05&	0.89	$\pm$0.05\\
 &	e3&	0.084	$\pm$0.002&	0.113	$\pm$0.003&	0.145	$\pm$0.003&	1.48	$\pm$0.04&	1.33	$\pm$0.04&	1.19	$\pm$0.03&	0.88	$\pm$0.06&	0.87	$\pm$0.06&	0.84	$\pm$0.06\\
 &	e4&	0.073	$\pm$0.002&	0.091	$\pm$0.002&	0.109	$\pm$0.003&	1.48	$\pm$0.04&	1.41	$\pm$0.04&	1.28	$\pm$0.04&	0.88	$\pm$0.06&	0.88	$\pm$0.06&	0.86	$\pm$0.06\\
 &	e5&	0.070	$\pm$0.002&	0.079	$\pm$0.002&	0.088	$\pm$0.002&	1.49	$\pm$0.04&	1.45	$\pm$0.04&	1.38	$\pm$0.04&	0.88	$\pm$0.06&	0.89	$\pm$0.06&	0.87	$\pm$0.05\\
 
 \hline
 
 500&	e2&	0.034	$\pm$0.001&	0.040	$\pm$0.001&	0.050	$\pm$0.001&	1.66	$\pm$0.05&	1.49	$\pm$0.04&	1.32	$\pm$0.04&	1.14	$\pm$0.06&	1.03	$\pm$0.05&	0.99	$\pm$0.07\\
 &	e3&	0.040	$\pm$0.001&	0.064	$\pm$0.002&	0.096	$\pm$0.002&	1.55	$\pm$0.05&	1.32	$\pm$0.03&	1.14	$\pm$0.03&	0.92	$\pm$0.06&	0.88	$\pm$0.05&	0.83	$\pm$0.06\\
 &	e4&	0.036	$\pm$0.001&	0.046	$\pm$0.001&	0.063	$\pm$0.001&	1.59	$\pm$0.05&	1.40	$\pm$0.04&	1.22	$\pm$0.03&	0.95	$\pm$0.07&	0.89	$\pm$0.05&	0.86	$\pm$0.06\\
 &	e5&	0.034	$\pm$0.001&	0.040	$\pm$0.001&	0.047	$\pm$0.001&	1.57	$\pm$0.04&	1.53	$\pm$0.05&	1.39	$\pm$0.04&	0.93	$\pm$0.06&	0.94	$\pm$0.06&	0.92	$\pm$0.06\\
 \hline

\end{tabular} }
\end{center}
\end{table*}